\documentclass[a4paper,11pt]{article}
\pdfoutput=1

\usepackage{ifthen}
\newboolean{pdflatex}
\setboolean{pdflatex}{true}
\newboolean{uprightparticles}
\setboolean{uprightparticles}{false}

\newboolean{inbibliography}
\setboolean{inbibliography}{false}

\usepackage{mciteplus}
\usepackage{enumitem}
\usepackage{jinstpub}

\usepackage[table]{xcolor}
\usepackage{multirow}

\title{\boldmath Application of machine learning techniques to lepton energy reconstruction in water Cherenkov detectors}

\author[a]{E. Drakopoulou$^{1,}$ \note{Corresponding author.},}
\author[a]{G. A. Cowan,}
\author[a]{M. D. Needham,}
\author[a]{S. Playfer and}
\author[a]{M. Taani}

\affiliation[a]{School of Physics and Astronomy, University of
  Edinburgh, Edinburgh, United Kingdom}

\emailAdd{edrakopo@staffmail.ed.ac.uk}

\abstract{The application of machine learning techniques to the reconstruction of lepton energies
    in water Cherenkov detectors is discussed and illustrated for TITUS, a proposed intermediate detector for the
Hyper-Kamiokande experiment. It is found that applying these
techniques leads to an improvement of more than 50\% in the energy resolution for all lepton energies compared to an approach based upon lookup tables. Machine learning techniques can be easily applied to different detector configurations and the results are comparable to likelihood-function based techniques that are currently used.}


\begin{document}
\maketitle
\flushbottom

\section{Introduction}
\label{intro}
In the last decades the use of machine learning techniques for
selection and reconstruction of events has become commonplace in
experimental particle physics. This is driven by
increasing computing power and the availability of powerful
open-source toolkits such as the ROOT-based TMVA framework~\cite{Hocker:2007ht} and Python's
Scikit-Learn toolkit~\cite{scikitlearn}. Current
approaches are based upon either variants of Boosted
Decision Trees (BDTs) or Neural Networks (NNs) with large numbers of hidden layers.

In this paper we investigate the application of machine learning
techniques for lepton ener-gy reconstruction in water Cherenkov detectors,
focussing on their use in the proposed TITUS intermediate detector~\cite{Andreopoulos:2016rqc} for the Hyper-Kamiokande experiment~\cite{Abe:2015zbg}.
The ideas discussed are generic and applicable to similar running or planned water Cherenkov experiments, such as Super-Kamiokande~\cite{superk}, ANNIE~\cite{WChSandBox},
E61~\cite{E61} and Hyper-Kamiokande itself. This paper is organised as follows. First, we describe the geometry of the
TITUS detector (Section~\ref{titus}), then we describe the input
variables that are used to train the multivariate methods
considered (Section~\ref{multivariate}) before presenting the results obtained
(Section~\ref{results}). Finally, we present the results achieved by other experiments (Section ~\ref{discussion}) and we discuss the systematic uncertainties and how they can affect the lepton energy reconstruction (Section~\ref{robustness}).

\section{The TITUS detector}
\label{titus}
TITUS~\cite{Andreopoulos:2016rqc} was a proposed intermediate-baseline detector for the next generation
long-baseline neutrino experiment, Hyper-Kamiokande\footnote{The TITUS and nuPRISM~\cite{nuprism} collaborations have recently merged to form a new collaboration E61.}.
By positioning a water Cherenkov detector at 1-2 km from the neutrino production target, systematic uncertainties due to the beam flux, which are crucial for studies of \textit{CP} violation
in the neutrino sector, can be reduced.

The proposed layout of TITUS is detailed in Ref.~\cite{Andreopoulos:2016rqc}. The detector is a horizontal cylinder 22 m long and 11 m in diameter filled with water, giving a total mass of 2.1 kton. In the baseline design the tank walls are covered by a total of 5131 12'' photomultiplier tubes (PMTs) leading to a photocoverage of $40 \%$. In the following a quantum efficiency of $22 \%$ is assumed. The TITUS geometry is well suited to test the performance of different machine learning techniques for energy reconstruction.

Charged-current neutrino interactions in water produce a charged lepton, accompanied by neutrons, protons and other particles. As the charged particles traverse the water volume they emit Cherenkov photons that
are collected by the PMTs, the information from which can subsequently be used to reconstruct the energy, position and direction of the charged lepton. The neutrino beam from the J-PARC proton accelerator facility in Japan has energy of $0.2-4$~GeV. Charged current quasi-elastic (CCQE) interactions lead to lepton energies of $0.06-3.6$~GeV in the TITUS detector. The initial design studies for the energy reconstruction of muons and electrons in TITUS was performed using a set of lookup
tables~\cite{Andreopoulos:2016rqc} that used information on the total number of photoelectrons which have been assigned to Cherenkov rings and the minimum perperdincular distance from the interaction vertex to the closest walls and endcaps of the tank~\cite{Andreopoulos:2016rqc}.

The neutrino interactions for this study are simulated with the NEUT v5.3.3 neutrino event generator~\cite{Andreopoulos:2016rqc}. The neutrino spectrum is set to that expected for the J-PARC beam. The detector simulation is performed using the WChSandBox~\cite{WChSandBox} simulation package which is based upon the Geant4~\cite{geant} toolkit. As well as simulating secondary interactions, the simulation models the expected PMT pulse size, shape, timing resolution and wavelength-dependent quantum efficiency. Simulated events are reconstructed as described in
Ref.~\cite{Andreopoulos:2016rqc}. Initially, combinations of PMT hits are used to obtain a solution for the position, the direction and the time of origin of an interaction. The energy of each event is approximated from the number of hits recorded by the detector and if it is above a threshold of 60 MeV, which was chosen to reject Michel electrons and neutron capture signals, the event is passed to a second reconstruction step. This step searches for Cherenkov rings and assigns observed photoelectrons
to the rings. The timing information of each hit is then used to improve the reconstructed vertex position and direction. Both the muon and electron hypotheses are considered using a lookup-based method to determine the energy and a likelihood-based particle identification method to determine whether the ring is electron-like or muon-like~\cite{Andreopoulos:2016rqc}. In the following only charged-current quasi-elastic neutrino interactions are used for the lepton energy reconstruction studies. This is a common practice which facilitates comparisons of the energy resolution between different algorithms~\cite{hyperk}. 

\section{Machine learning for energy regression}
\label{multivariate}
A wide range of machine learning techniques is available in open source packages. For this study a subset of techniques has been considered that show good performance and are easy and fast to train. These are the gradient BDT algorithms available in the Scikit-Learn~\cite{scikitlearn} (Scikit-Learn 0.18.2)  and TMVA packages~\cite{Hocker:2007ht} (ROOT 5.34/23), a multi-layer percepton NN (MLPNN) from the TMVA package and a multi-layer NN implemented using TensorFlow (TNN)~\cite{tensorflow2015-whitepaper} via the Keras library in Python (Keras 1.2.2)~\cite{keras}. Other techniques that have been tested but are not presented here include k-Nearest Neighbour estimator, Linear Discriminant analysis, Likelihood Estimator using self-adapting phase-space binning (PDE-Foam) and Multidimensional Likelihood estimator (PDE range-search approach). Multidimensional Likelihood and k-Nearest Neighbour estimators show comparable performance to NN algorithms that are presented here but are slower to train. To avoid complexity we do not present these results. Linear Discriminant analysis and PDE-Foam do not show good performance. 
 
The methods considered were applied to both muons and electrons produced by $\nu_{\mu}$ and $\nu_{e}$ events and comparable performance was found for both lepton flavours.
The multivariate algorithms are trained using reconstructed track and interaction vertex information to predict the energy of the charged lepton. Specifically, the input variables are:
\begin{description}
\item[$N_{\rm C}^{\rm Hit}$]: Clusters of hits in time coincidence are searched for and the total number of hits assigned to clusters is calculated~\cite{Andreopoulos:2016rqc}. In particular, the photoelectrons recorded by the PMTs are clustered according to time coincidences and the total number of selected hits in clusters is measured. Hits in neighbouring PMTs with spacing of less than10~ns are grouped in clusters and each cluster allows for a maximum spacing of 200~ns between the first and last hit of the cluster.
\item[$N_{\rm Ring}^{\rm Hit}$]: The number of assigned hits from the Cherenkov rings~\cite{Andreopoulos:2016rqc}. The total number of photoelectrons assigned to all Cherenkov rings is measured. In the TITUS reconstruction a Hough transform, similar to that used in Super Kamiokande, is employed to search for rings. More details can be found in~\cite{Andreopoulos:2016rqc} and~\cite{superKreco}. 
\item[$D_{\rm R}^{\rm Wall}$]: The minimum distance from the reconstructed vertex to the closest barrel wall.
\item[$D_{\rm Z}^{\rm Wall}$]: The distance from the reconstructed vertex to the nearest detector endcap.
\item[$L$]: The track length estimated from the observed pattern of PMT hits. The track length is calculated as the distance between the reconstructed vertex and the last photon emission point under the hypothesis of Cherenkov photons being produced along the track.
\end{description}

\begin{figure}[t]
\centering
  \includegraphics[width=1.\linewidth]{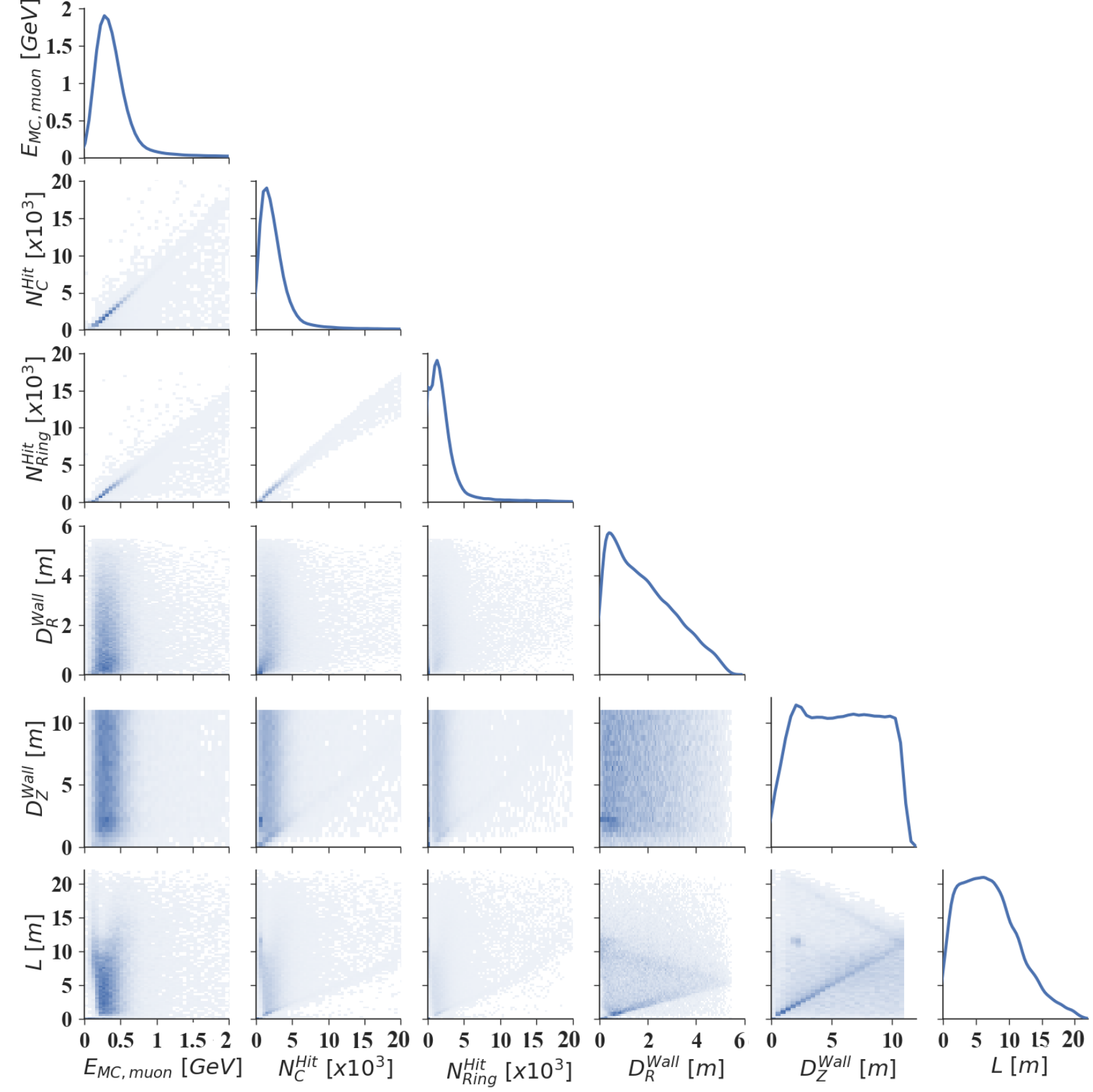}
  \caption{Two-dimensional distributions of event observables that are used as input to the machine learning algorithms, the simulated charged-lepton energy ($E_{\rm MC}$) and their scatter plots showing the correlations between observables.} 
\label{plot_Vars}
\end{figure}

\begin{figure}[t]
\centering
  \includegraphics[width=0.7\linewidth]{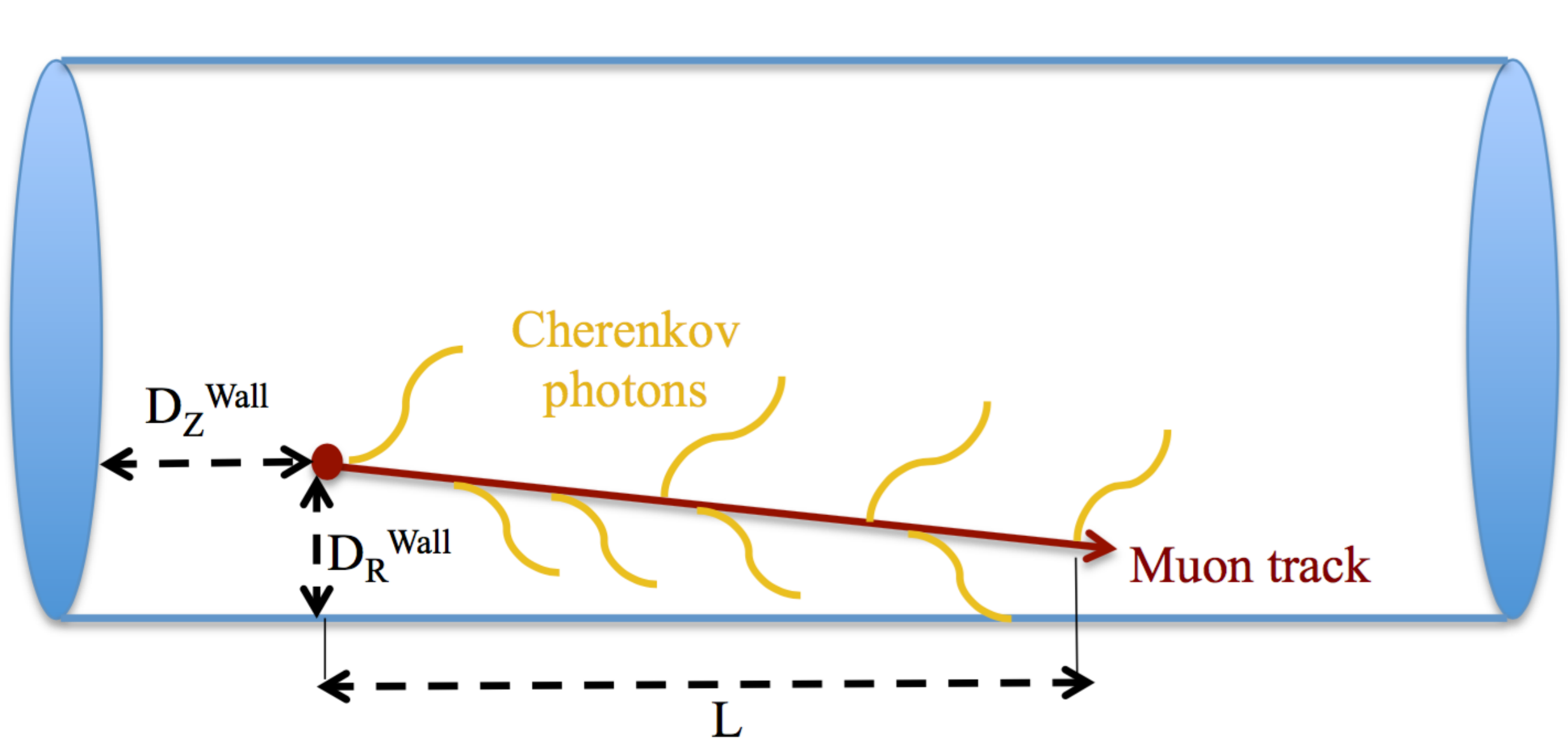}
 \caption{Variables $D_{\rm R}^{\rm Wall}$, $D_{\rm Z}^{\rm Wall}$ and $L$ account for the distance of the muon track from the detector walls. This illustration does not represent the detector scale. Only a few representative Cherenkov photons are shown.}
 \label{fig:wallplot}
\end{figure}

Figure~\ref{plot_Vars} shows the distributions of the input variables and the simulated charged-lepton energy that we are attempting to reconstruct. The variables $N_{\rm C}^{\rm Hit}$ and $N_{\rm Ring}^{\rm Hit}$ clearly have a strong dependence on the charged-lepton energy. The variables $D_{\rm R}^{\rm Wall}$, $D_{\rm Z}^{\rm Wall}$ and $L$ (Figure~\ref{fig:wallplot}) account for the event topology and improve the performance for events that are only partially contained within the detector volume. Removing these variables degrades the energy resolution. Different input variables including the number of PMTs with hits and the distances from the reconstructed vertex to the downstream barrel walls or endcaps were also tested but were not improving the energy reconstruction and were removed. It was observed that removing more variables from $N_{\rm C}^{\rm Hit}$, $N_{\rm Ring}^{\rm Hit}$, $D_{\rm R}^{\rm Wall}$, $D_{\rm Z}^{\rm Wall}$, $L$ degrades the performance. The input variables used for the machine learning methods and for the lookup tables are reported in Table ~\ref{table:a}.

\begin{table}[t]
\caption{The input variables used for each method. } 
\vspace{\baselineskip} 
 \centering
   \begin{tabular}[b]{|c | c | c | c| c| c| }
     \hline
     \multirow{2}{*}{\textbf{Method}} & \multicolumn{5}{c}{\textbf{${Input Variables}$}} \vline \\  \cline{2-6}
      & \textbf{$N_{\rm C}^{\rm Hit}$}  & \textbf{$N_{\rm Ring}^{\rm Hit}$} & \textbf{$D_{\rm R}^{\rm Wall}$} & \textbf{$D_{\rm Z}^{\rm Wall}$} & \textbf{$L$} \\  \hline
     Scikit-Learn BDT & $\bullet$ & $\bullet$ & $\bullet$ & $\bullet$ & $\bullet$ \\
    TMVA BDT & $\bullet$ & $\bullet$ & $\bullet$ & $\bullet$ & $\bullet$  \\
    Tensorflow NN & $\bullet$ & $\bullet$ & $\bullet$ & $\bullet$ & $\bullet$  \\
    TMVA  NN & $\bullet$ & $\bullet$ & $\bullet$ & $\bullet$ & $\bullet$  \\
    Lookup Tables & $\bullet$ & $\bullet$ & $\bullet$ & $\bullet$ &   \\
     \hline
   \end{tabular}
\label{table:a}
\end{table}

\begin{figure}[t]
\centering
  \includegraphics[width=0.49\linewidth]{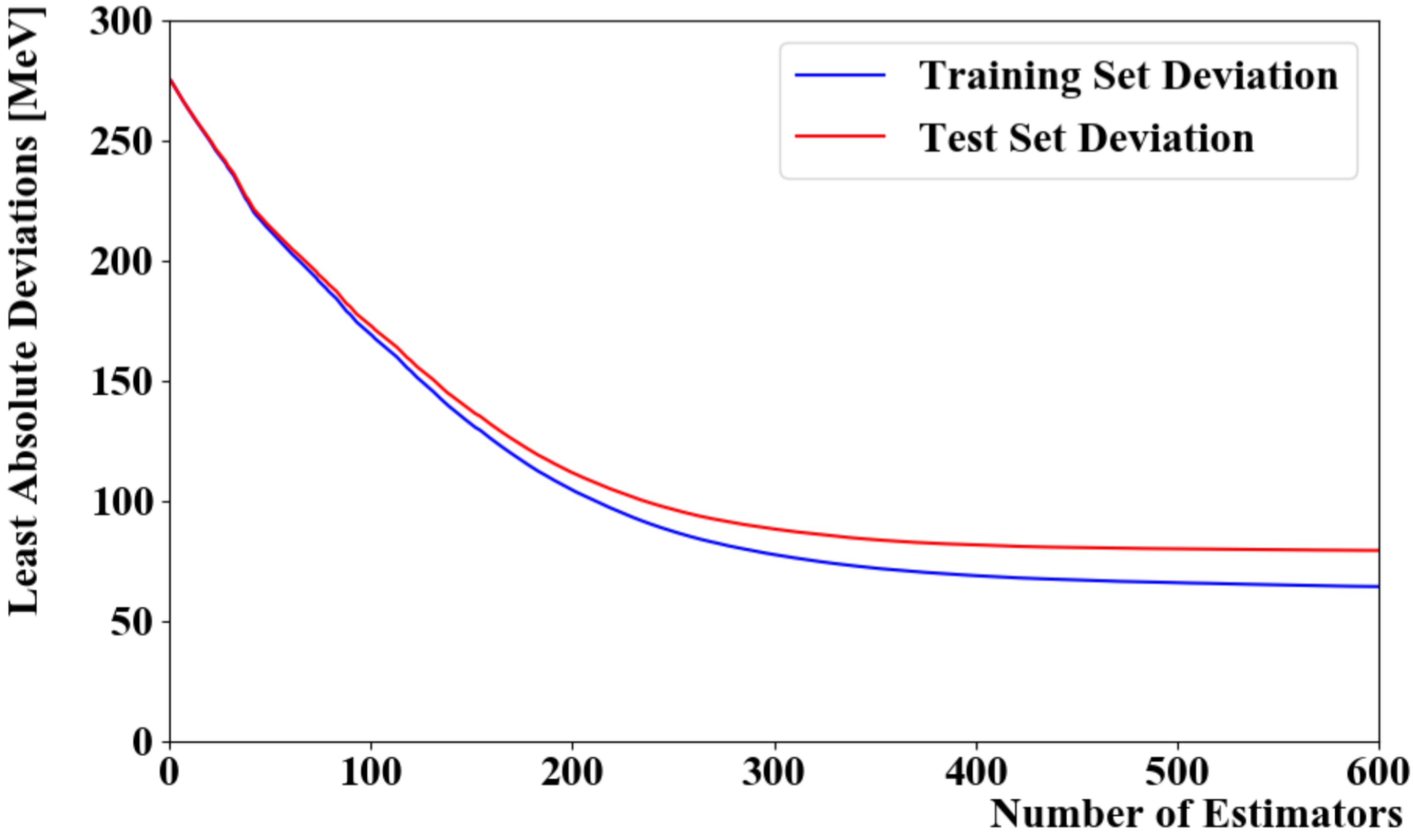}
  \caption{Evolution of the least absolute deviations of the charged-lepton energy (for simulated muon events)
  during the boosting steps ('Number of estimator') for the test and training samples for the Scikit-Learn BDT.}
\label{fig:train_tets}
\end{figure}

The distributions of all input variables are rescaled so their maximal values are unity (and their minimum are zero) before they are passed to the machine learning algorithms. This is a common practice in machine learning algorithms so that the input variables get a corresponding value to the activation function. For training, a sample of 100,000 simulated events generated with the lepton energy spectrum corresponding to the neutrino beam from J-PARC for CCQE events in the TITUS detector is used. To check for overtraining the dataset is divided into test (30\% of the MC sample) and training (70\% of the MC sample) samples. As an example of the training phase, Figure~\ref{fig:train_tets} shows the convergence of the Scikit-Learn BDT using the least absolute deviation estimator, which is one of the available options in Scikit-Learn package. This minimises the sum of the absolute values of the residuals between the true and the predicted value~\cite{LAD2} and compared to a least squares approach, is less sensitive to outliers~\cite{LAD1}.

All the algorithms considered have many adjustable parameters. For the Scikit-Learn and Tensorflow package these parameters are optimised using the rectangular grid search (GridSearchCV) method available in the Scikit-Learn package~\cite{scikitlearn}. This works as follows. Keeping all other parameters fixed the method estimates the least absolute deviations for all the available options of the parameter that needs to be adjusted. The method returns the parameter setting that gives the lowest deviation between the true and the predicted value. The procedure is repeated for all adjustable parameters to find the combination that gives the best results.

For the optimisation, leptons with energy in the range 200-600~MeV are used. In this way the best Scikit-Learn BDT performance was found to be for maximum tree depth of 10 for an individual regression of 600 estimators with a learning rate of 0.01. In the case of the TNN method, this led to five neurons in the hidden layer. The rectifier activation function was chosen and the sample was trained for 10 epochs with a batch size of 10 using the Nesterov Adam optimiser~\cite{NAO}
with a learning rate of 0.1. The rectangular grid search is not available in the TMVA package. Instead, several combinations were studied to find one that gives the best performance. The optimised TMVA BDT has 10000 trees with a maximum depth of three, a learning rate of 0.1; all other parameters set to the default values. The optimised MLPNN method has one hidden layer with 14 nodes and 1500 training cycles. The chosen activation function was the hyperbolic tangent  and the training method was the Broyden-Fletcher-Goldfarb-Shannon (BFGS) method~\cite{Hocker:2007ht}.


\section{Results}
\label{results}

After training, the performance of each method was evaluated using an independent simulation sample. 
To evaluate the performance the fractional energy residual is studied. It is defined as: 
\begin{equation*}
\frac{\Delta E}{E} = \frac{E^{\mathrm{rec}} - E^{\mathrm{MC}}}{E^{\mathrm{MC}}}
\end{equation*}
where $E^{\mathrm{MC}}$ and $E^{\mathrm{rec}}$ are the true and reconstructed charged-lepton
energy respectively.

\begin{figure}[t]
\centering
 \centering
   \includegraphics[width=0.49\linewidth]{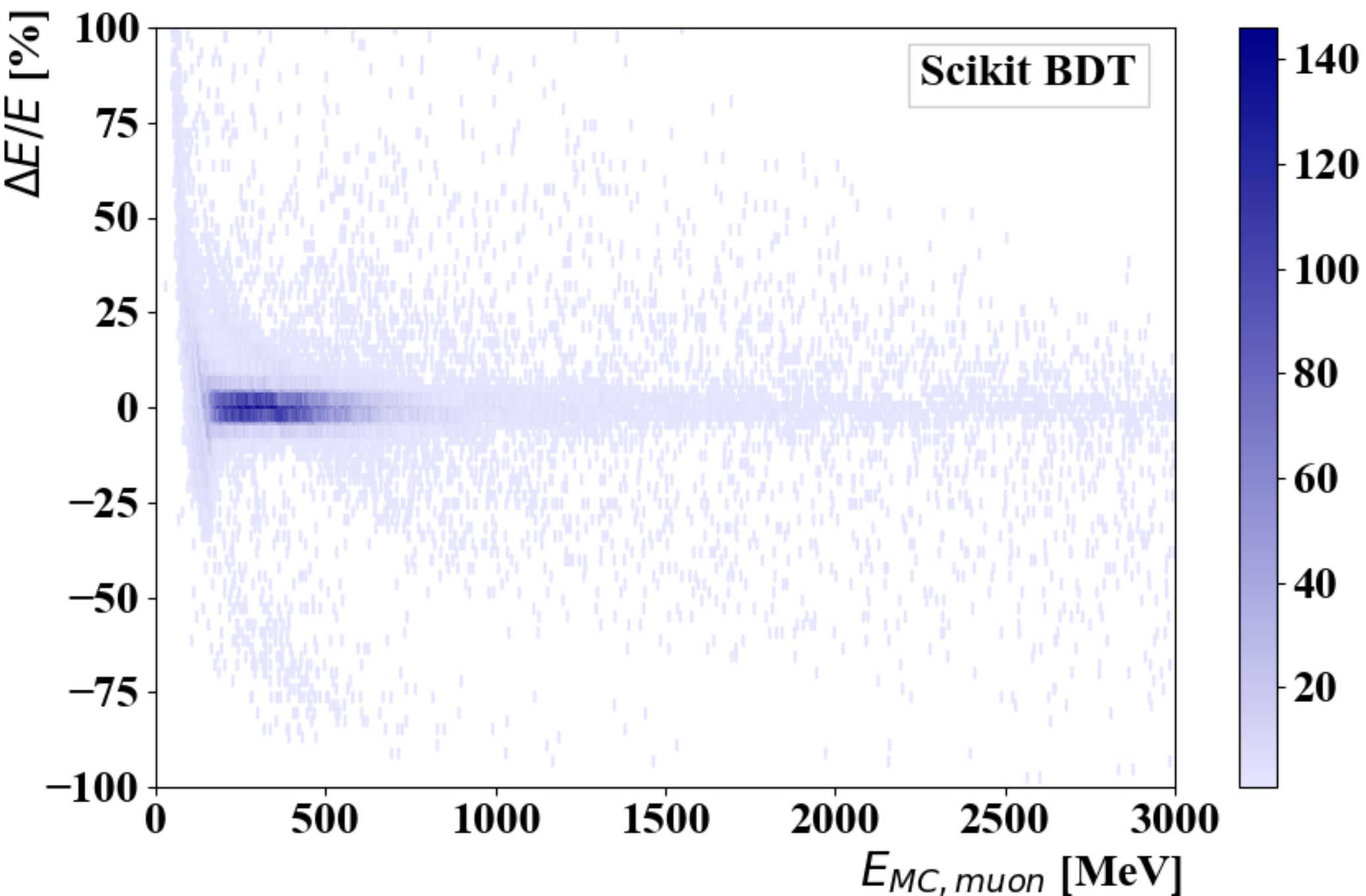}
 \centering
  \includegraphics[width=0.49\linewidth]{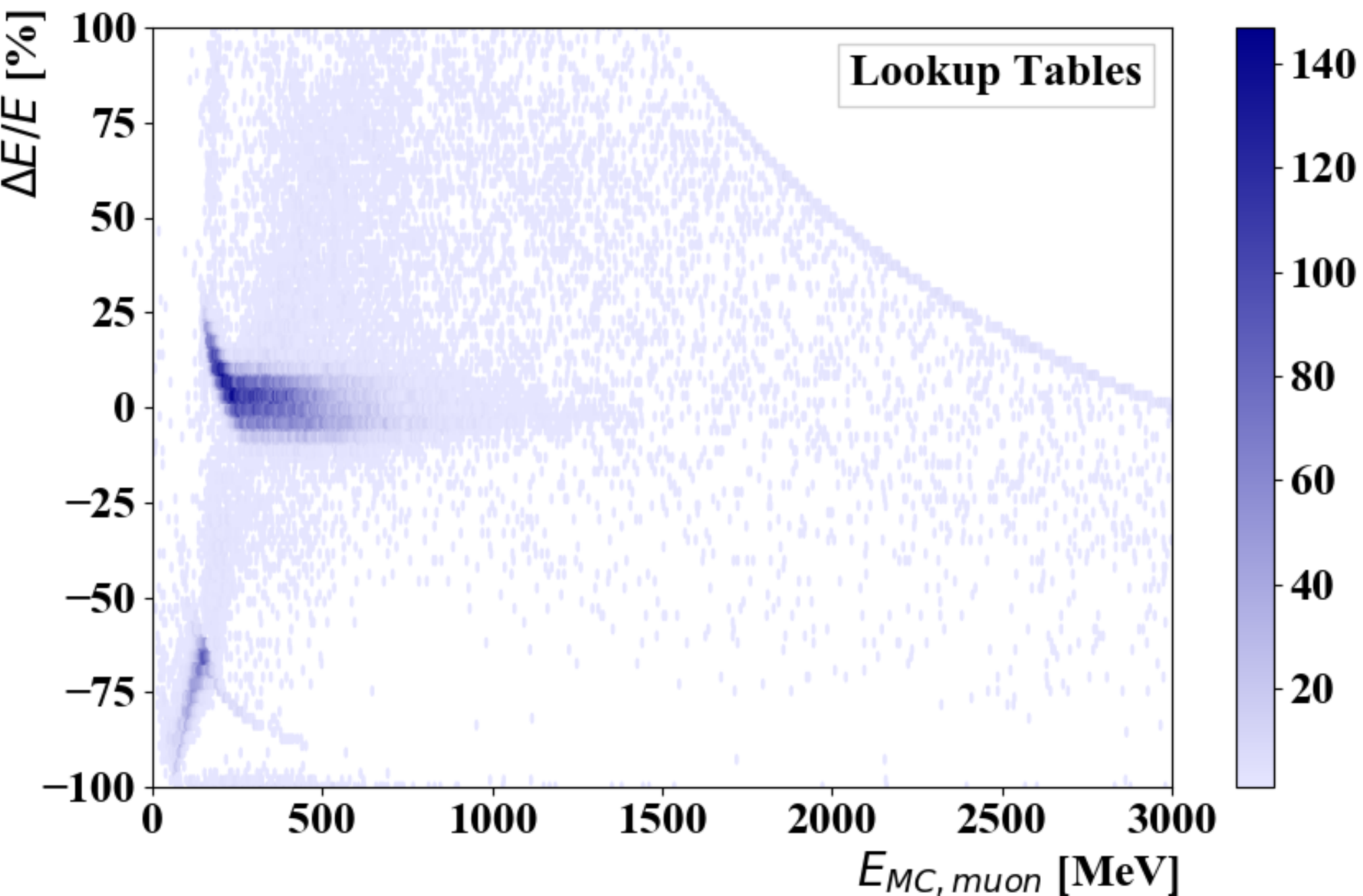}
\caption{The ${\Delta E}/{E}$ distribution as a function of the MC muon
  energy for the Scikit-Learn BDT (left) and the looukup tables (right) for events in the fiducial volume. The colour code represents the density of events for each bin.
}
\label{fig:resINfid}
\end{figure}

In the analysis of neutrino interactions in water Cherenkov detectors
it is customary to define a fiducial volume. For TITUS this is
a cylindrical volume 1~m away from each detector wall. Figure~\ref{fig:resINfid} shows ${\Delta E}/{E}$ versus
$E^{\mathrm{MC}}$ for the  Scikit-Learn BDT (left) and the lookup tables (right). Compared to the lookup-table based approach,
the BDT performs better across the full energy
range. The Cherenkov threshold for muons is 160~MeV and muons below this energy are not reconstructed.

\begin{figure}[t]
\centering
 \includegraphics[width=0.49\linewidth]{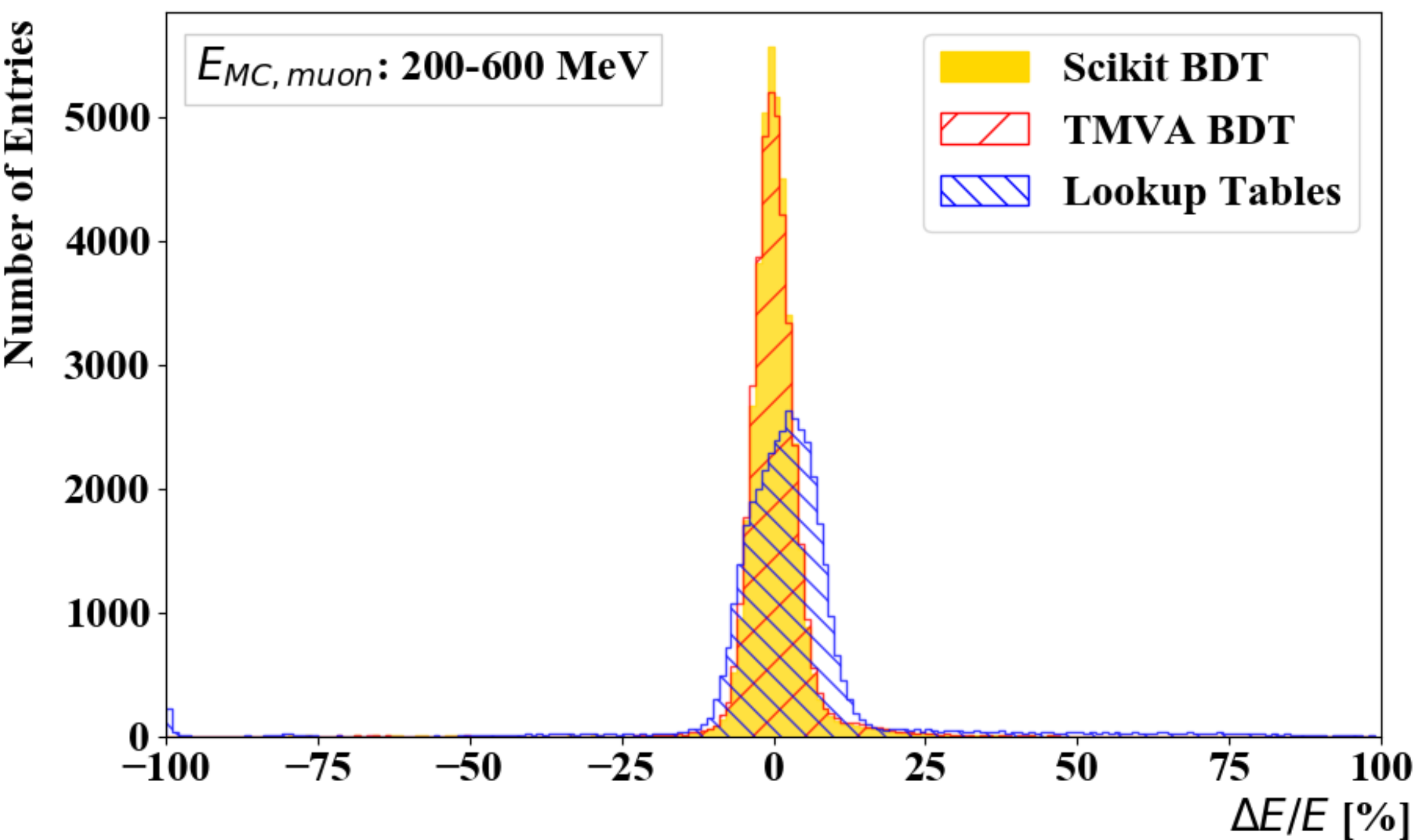}
 \centering
 \includegraphics[width=0.49\linewidth]{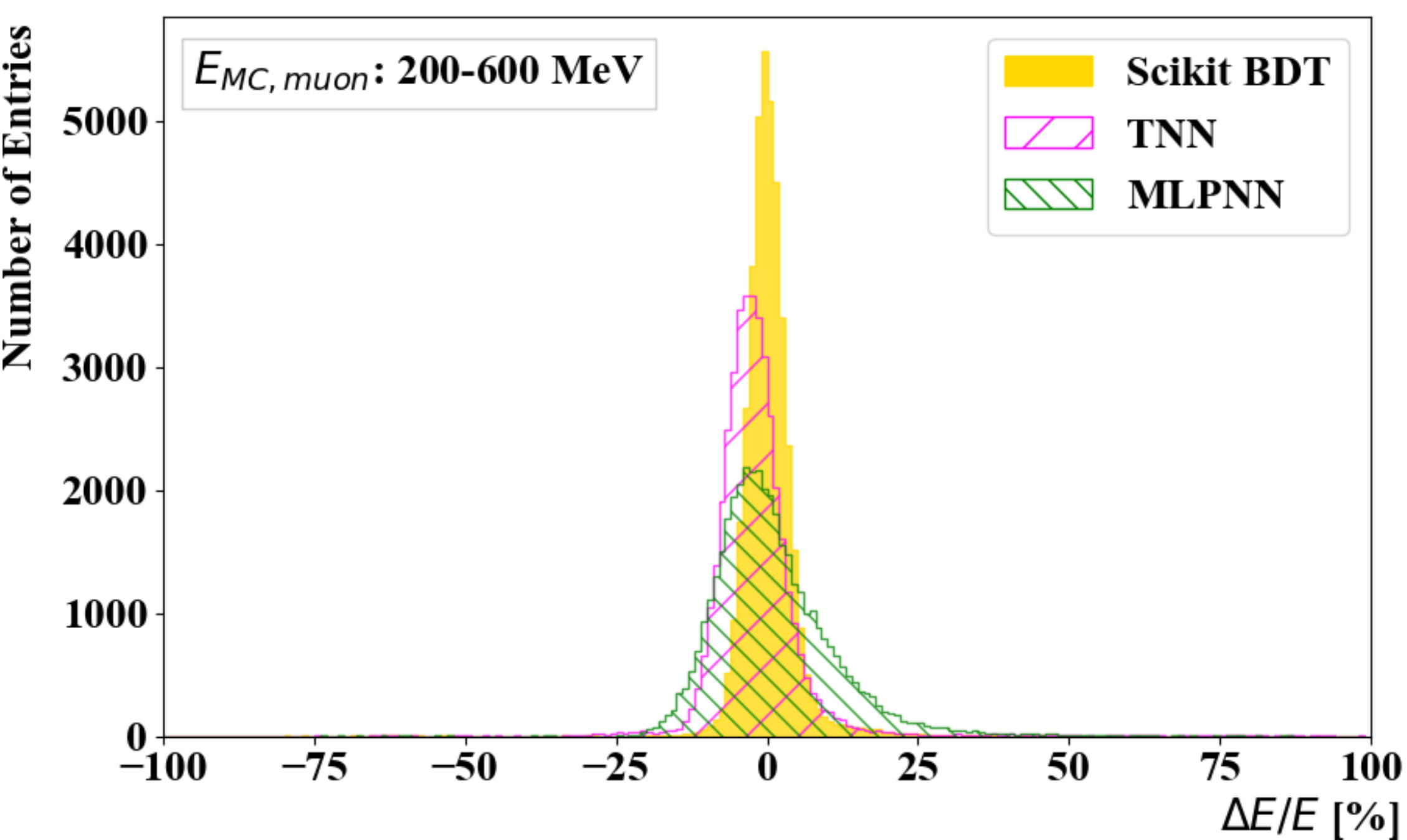}
  \caption{The distribution of ${\Delta E}/{E}$ for events in the fiducial volume and muon energies from 200-600 MeV for the different machine learning algorithms.}
  \label{fig:res200600INFID}
\end{figure}

\begin{figure}[t]
\centering
 \includegraphics[width=0.49\linewidth]{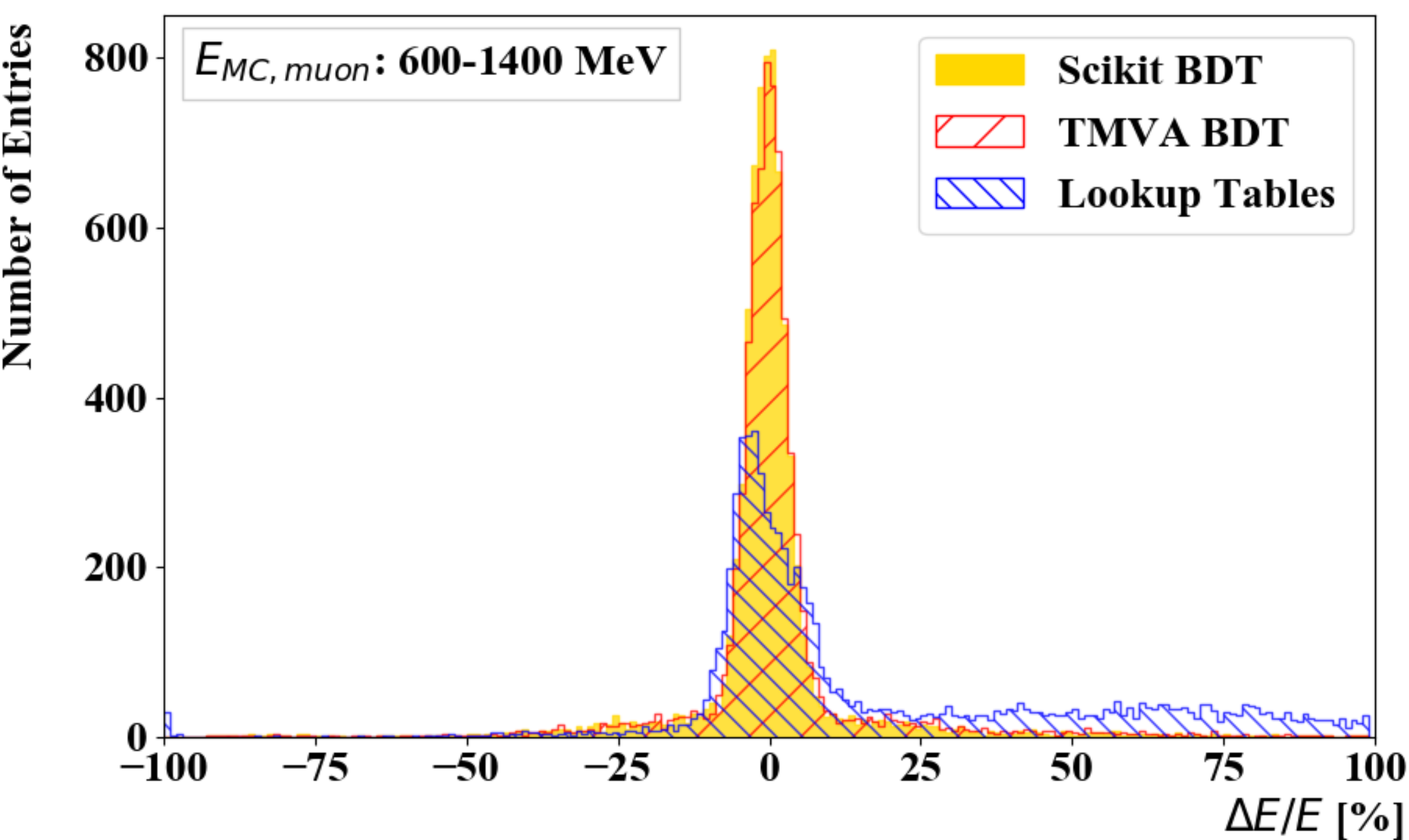}
 \centering
 \includegraphics[width=0.49\linewidth]{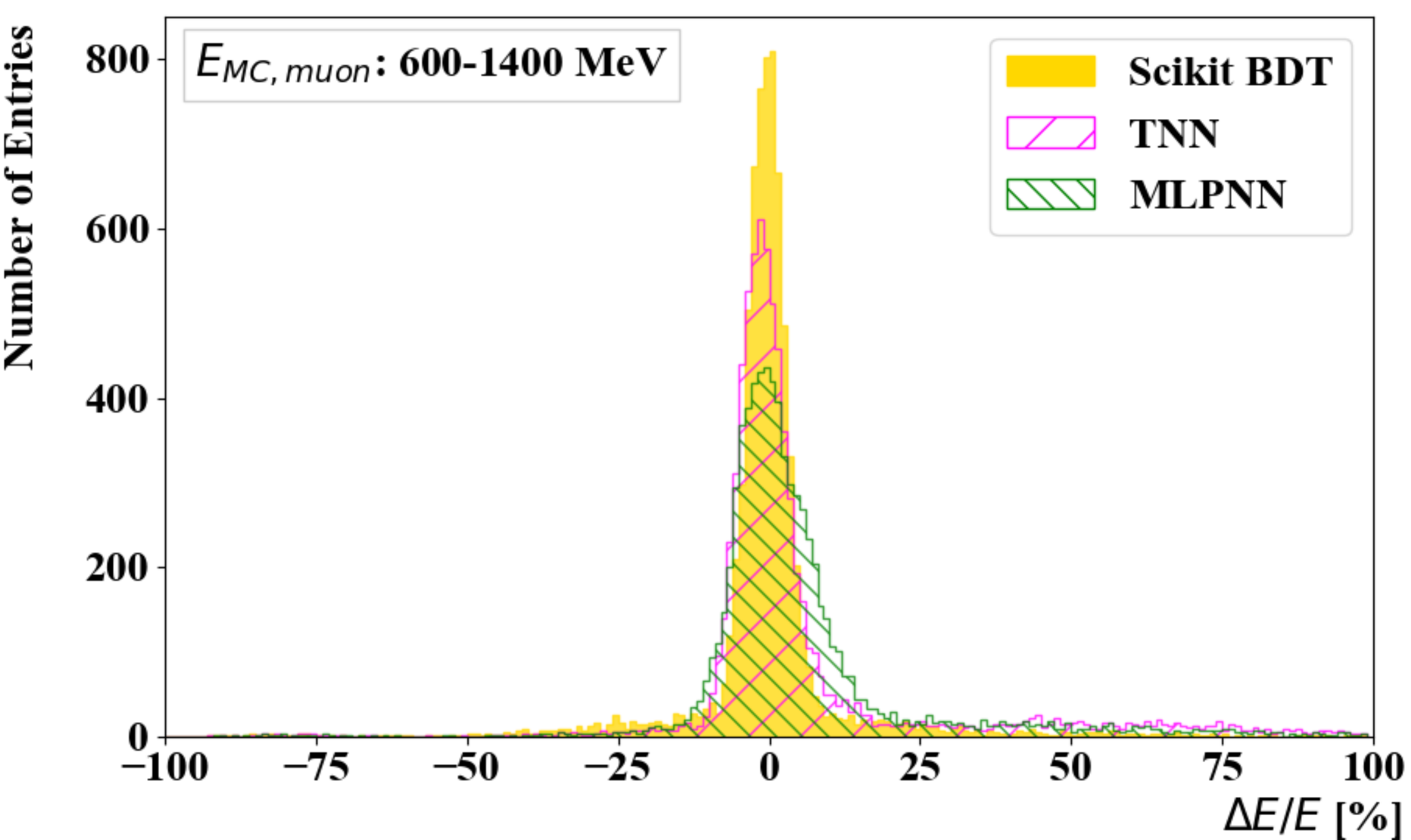}
  \caption{The distribution of ${\Delta E}/{E}$ for events in the fiducial volume and muon energies from 600-1400 MeV for the different machine learning algorithms.}
  \label{fig:res200600INFID_HE}
\end{figure}

The distribution of ${\Delta E}/{E}$ for the various methods considered is shown in Figure~\ref{fig:res200600INFID} for muon energies from 200-600 MeV, which is the energy region with the largest event yield. The corresponding distributions for muons with higher energies, from 600-1400 MeV, are shown in Figure~\ref{fig:res200600INFID_HE}. 


To quantify the performance of each method the following approach was
used. First, a bifurcated Gaussian is fitted to the core of the
distribution defined as $|{\Delta E}/{E}| < 0.1$. This fit gives the
mean $\mu$ and the core resolution of the left and right side of the
distribution ($\sigma_L$ and $\sigma_R$). To give an estimate of the
non-Gaussian tail, the fraction of events outside the range $\mu -
3\sigma_L$ to $\mu + 3\sigma_R$ is calculated. The results of these
fits are summarised in Table~\ref{table:ta} for muon energies from 200-600 MeV
and in Table~\ref{table:ta1} for muon energies from 600-1400 MeV.
The best resolution and smallest tail-fractions are seen for the two BDT-based
approaches, which significantly outperform the lookup tables and give better results than the Neural Networks. The TNN
method also gives reasonable performance with a lower percentage of events in the tails compared to the lookup tables.
However, this method underestimates the muon energy resulting in $\mu \simeq$ $-$3.80\% ($-$2.14\%) for muon energies from 200-600 MeV (from 600-1400 MeV). Muons at high energies (from 600-1400 MeV) can escape the instrumented volume leading to a larger percentage of events in the tails compared to muons with lower energies (from 200-600 MeV). The use of Scikit-Learn and TMVA BDTs leads to a smaller underestimation of $\mu \simeq$ $-$0.48\% and $-$0.54\% ($-$0.68\% and $-$0.54\%) for muon energies from 200-600 MeV (from 600-1400 MeV) respectively.

\begin{table}[t]
\caption{Fitted muon energy resolution parameters for muon neutrino interactions within the fiducial volume with the muon energy in the range 200-600 MeV. The quoted uncertainties are statistical only.} 
\vspace{\baselineskip} 
 \centering
   \begin{tabular}[b]{|c | c | c | c| c|}
     \hline
     \multirow{2}{*}{\textbf{Method}} & \multicolumn{4}{c}{\textbf{${\Delta E}/{E}~\%$}} \vline \\  \cline{2-5}
      & \textbf{$\mu$}  & \textbf{$\sigma_R$} & \textbf{$\sigma_L$} & Tail$~\%$ \\  \hline
     Scikit-Learn BDT & {$-$}0.48$\pm$0.04  & 3.39$\pm$0.03 & 2.86$\pm$0.02 &4.49 \\
     TMVA BDT & {$-$}0.54$\pm$0.04 &3.53$\pm$0.03  &2.95$\pm$0.03 &5.08 \\
     Tensorflow NN & {$-$}3.80$\pm$0.10  & 5.28$\pm$0.06 & 3.52$\pm$0.09 & 6.04\\
     TMVA  NN & {$-$}4.32$\pm$0.14  & 9.42$\pm$0.16 & 4.26$\pm$0.16 & 5.04\\
     Lookup Tables & \phantom{$-$}2.39$\pm$0.16 & 5.30$\pm$0.18 & 6.28$\pm$0.12  &9.46 \\
     \hline
   \end{tabular}
\label{table:ta}
\end{table}

\begin{table}[t]
 \caption{Fitted energy resolution parameters for muon neutrino interactions within the fiducial volume with the muon energy in the range 600-1400 MeV. The quoted uncertainties are statistical only.}
 \vspace{\baselineskip}
 \centering
   \begin{tabular}[b]{|c | c | c | c| c|}
     \hline
     \multirow{2}{*}{\textbf{Method}} & \multicolumn{4}{c}{\textbf{${\Delta E}/{E}~\%$}} \vline \\  \cline{2-5}
      & \textbf{$\mu$}  & \textbf{$\sigma_R$} & \textbf{$\sigma_L$} & Tail$~\%$ \\  \hline
     Scikit-Learn BDT & $-$0.68$\pm$0.09  &3.46$\pm$0.07  &3.11$\pm$0.06 &14.84 \\
     TMVA  BDT & $-$0.54$\pm$0.11 &3.64$\pm$0.07  &3.24$\pm$0.07  &15.90 \\
     Tensorflow NN & $-$2.14$\pm$0.16  &4.97$\pm$0.12  & 3.31$\pm$0.12 &18.94 \\
     TMVA  NN & $-$2.09$\pm$0.29  &7.68$\pm$0.32  &3.97$\pm$0.22  &14.07 \\
     Lookup Tables & $-$4.06$\pm$0.27 & 8.13$\pm$0.29 & 2.99$\pm$0.21  &37.08 \\
     \hline
   \end{tabular}
\label{table:ta1}
\end{table}

The improved muon energy reconstruction allows a larger fiducial volume to be used. The effect of extending the fiducial requirement has also been studied. The 1~m distance of the reconstructed vertex from the detector wall is relaxed to: 0.8~m distance from the detector cylindrical surface and 0.9~m distance from the endcaps. These requirements define a new fiducial volume which is $\sim$ 10~\% larger than the initial one.

\begin{figure}[t]
\centering
 \includegraphics[width=0.49\linewidth]{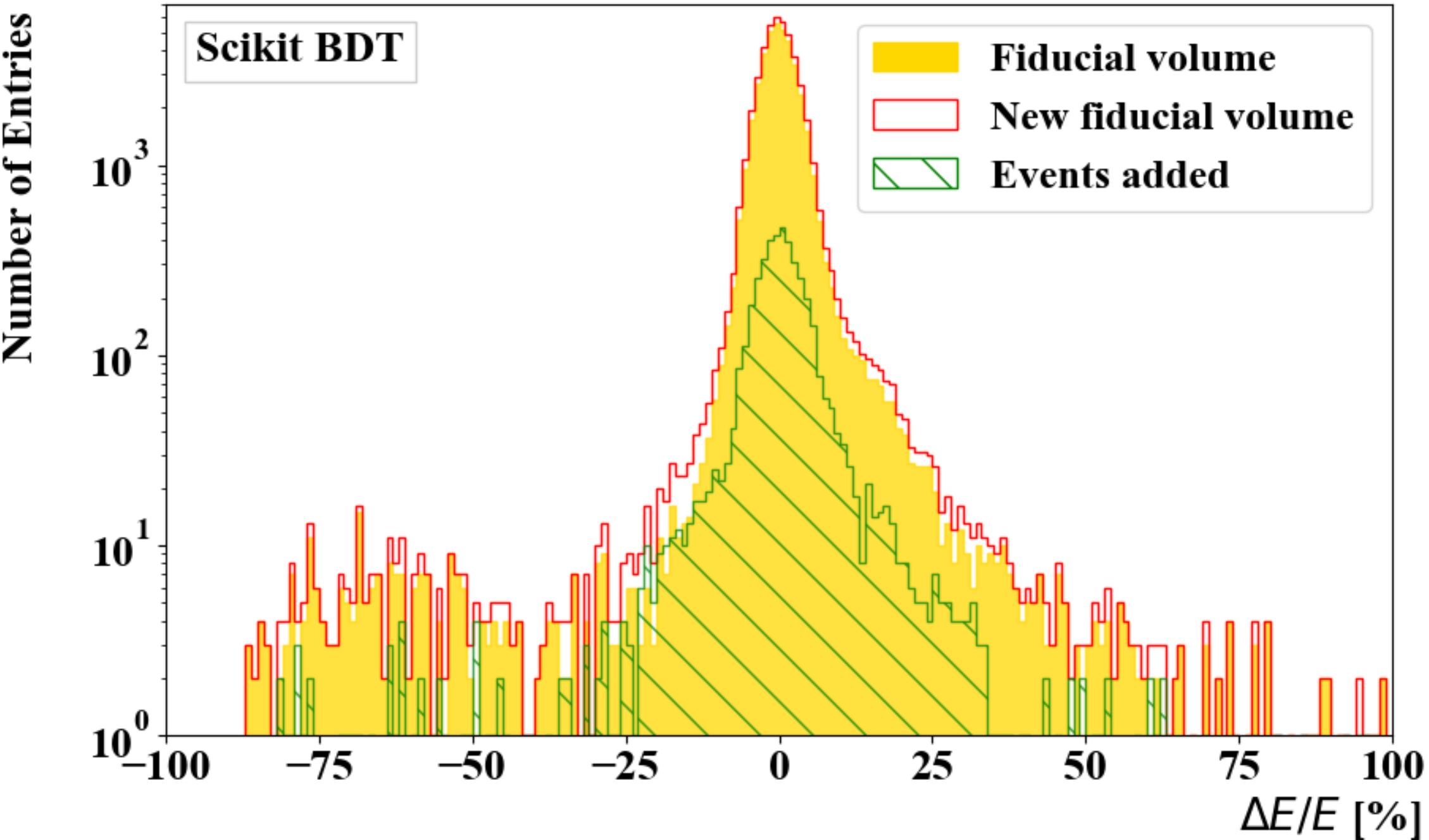}
  \centering
 \includegraphics[width=0.49\linewidth]{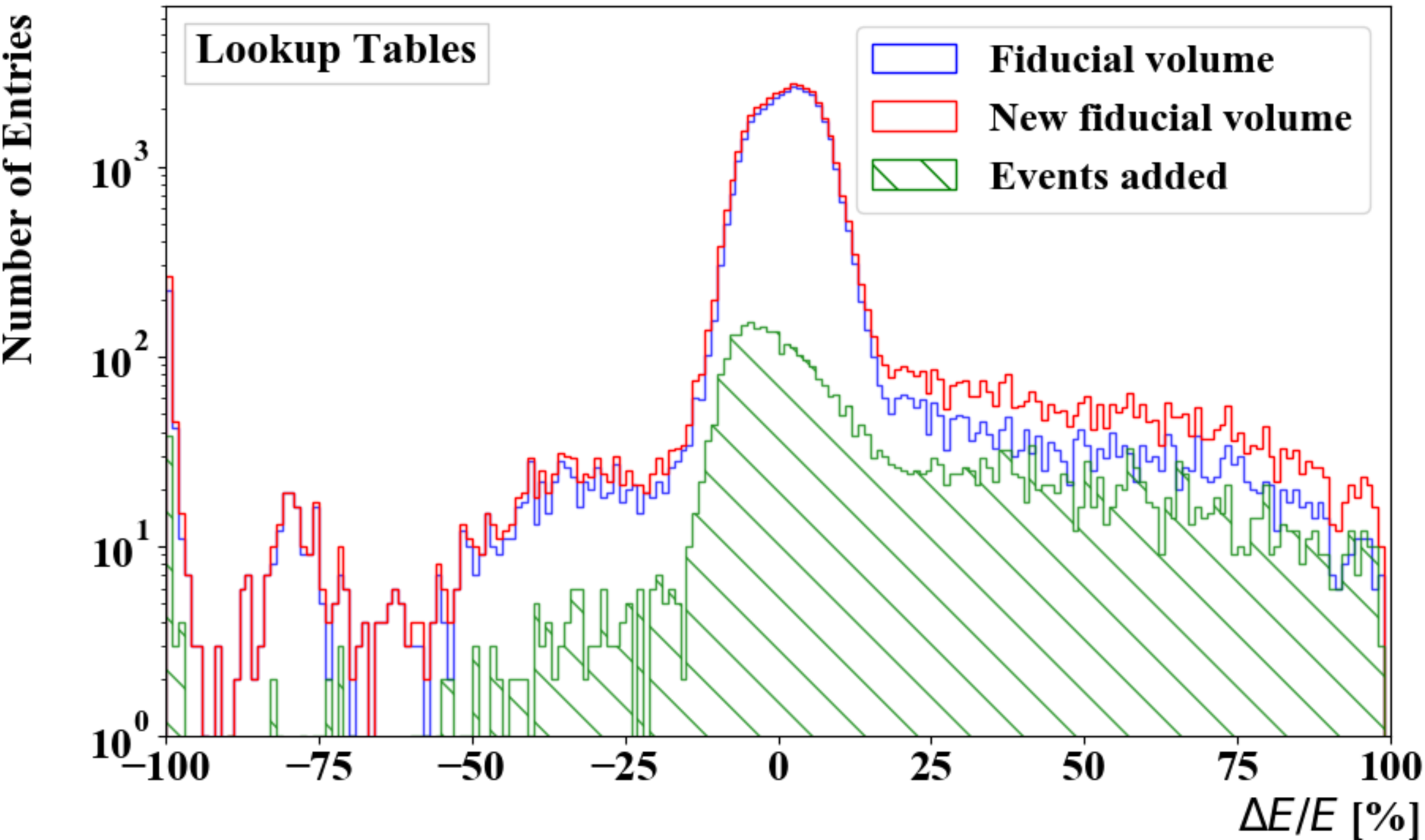}
  \caption{The distribution of ${\Delta E}/{E}$ with the Scikit-Learn BDT (left) and the lookup tables (right) for muons in the extended fiducial volume (red line), muons in the default fiducial volume and muons in the new fiducial volume, which do not satisfy the default fiducial volume selection (green line). The muons energies range from 200-600 MeV.}
  \label{additional_evts}
\end{figure}

The results of the fit procedure are reported in Table~\ref{table:tb}
for the muon energy range 200-600~MeV and in Table~\ref{table:tb1}
for 600-1400~MeV respectively. The machine learning-based
methods lead to better results than the lookup tables in particular
reducing the non-Gaussian tail. Both BDTs have comparable performance
and perform better than the NNs. The energy resolution with BDTs for
events in this extended fiducial volume is comparable to that in the nominal one. As an example, the distribution of the fractional energy residuals for the additional events in the extended fiducial volume for muon energies from 200-600 MeV is shown in Figure~\ref{additional_evts} for the Scikit-Learn BDT and the lookup tables.
Compared to the various considered methods the Scikit-Learn BDT gives the best performance.

\begin{table}[t]
\caption{Fitted energy resolution parameters for events in an extended fiducial volume and muon energies from 200-600 MeV for each method. The quoted uncertainties are statistical only.}
\vspace{\baselineskip}
 \centering
   \begin{tabular}[b]{|c | c | c | c| c|}
     \hline
     \multirow{2}{*}{\textbf{Method}} & \multicolumn{4}{c}{\textbf{${\Delta E}/{E}~\%$}} \vline \\  \cline{2-5}
      & \textbf{$\mu$}  & \textbf{$\sigma_R$} & \textbf{$\sigma_L$} & Tail$~\%$ \\  \hline
     Scikit-Learn BDT & {$-$}0.46$\pm$0.04 & 3.46$\pm$0.03 & 2.92$\pm$0.02 & 5.13\\
     TMVA  BDT & {$-$}0.52$\pm$0.04  & 3.61$\pm$0.03 &3.02$\pm$0.02  &5.74 \\
     Tensorflow NN & {$-$}3.98$\pm$0.09  & 5.44$\pm$0.06 & 3.49$\pm$0.08 & 6.94\\
     TMVA  NN& {$-$}4.35$\pm$0.14 & 9.75$\pm$0.16 & 4.22$\pm$0.16 & 5.85\\
     Lookup Tables & \phantom{$-$}2.31$\pm$0.18 & 5.39$\pm$0.19& 6.50$\pm$0.13 &12.30\\
     \hline
   \end{tabular}
\label{table:tb}
\end{table}

\begin{table}[t]
\caption{Fitted energy resolution parameters for events in an extended fiducial volume and muon energies from 600-1400 MeV for each method. The quoted uncertainties are statistical only.}
\vspace{\baselineskip}
 \centering
   \begin{tabular}[b]{|c | c | c | c| c|}
     \hline
    \multirow{2}{*}{\textbf{Method}} & \multicolumn{4}{c}{\textbf{${\Delta E}/{E}~\%$}} \vline \\  \cline{2-5}
      & \textbf{$\mu$}  & \textbf{$\sigma_R$} & \textbf{$\sigma_L$} & Tail$~\%$ \\  \hline
     Scikit-Learn BDT & $-$0.70$\pm$0.09  & 3.53$\pm$0.06 & 3.13$\pm$0.06 & 16.50\\
     TMVA  BDT & $-$0.52$\pm$0.10 & 3.70$\pm$0.07 & 3.28$\pm$0.07 & 17.39\\
     Tensorflow NN & $-$2.23$\pm$0.15  & 5.24$\pm$0.12 & 3.32$\pm$0.11 & 20.25\\
     TMVA  NN & $-$1.93$\pm$0.34 & 7.92$\pm$0.38 & 4.03$\pm$0.25 & 15.03\\
     Lookup Tables & $-$4.20$\pm$0.23 & 8.25$\pm$0.27 & 3.00$\pm$0.19  & 40.01\\
     \hline
   \end{tabular}
\label{table:tb1}
\end{table}

The better performance of the machine-learning based techniques, especially for events within the extended fiducial volume, is understood as follows. The events in the tail of the ${\Delta E}/{E}$ distribution with the lookup tables are close to the wall of the cylinder. Though this is the case, if the muon energy is high and it travels parallel to the wall, the track length can be long and a sizeable amount of energy is deposited. The lookup table approach assigns a lower energy in this case. The machine-learning techniques successfully identify this topology. Better results with the lookup tables could be obtained by taking this into account at the expense of increased dimensionality. In contrast, the BDT approach automatically learns how to parametrise phase space. In order to allow for a direct comparison between the Scikit-Learn BDT and the lookup table we trained and tested a BDT without the track length information (with $N_{\rm C}^{\rm Hit}$, $N_{\rm Ring}^{\rm Hit}$, $D_{\rm R}^{\rm Wall}$, $D_{\rm Z}^{\rm Wall}$) using variables also used by the lookup tables. The BDT outperforms the lookup-table based approach resulting in $\mu \simeq$ 0.31$\pm$0.09\%, $\sigma_R$=4.61$\pm$0.07\% and $\sigma_L$=5.21$\pm$0.07\% with 2.97\% of the events being in the tail of the distribution for muon energies from 200-600 MeV in the default fiducial volume. The corresponding values for the lookup table are: $\mu \simeq$ 2.39$\pm$0.16\%, $\sigma_R$= 5.30$\pm$0.18\% and  $\sigma_L$=6.28$\pm$0.12\% with 9.46\% of the events being in the tail of the distribution (Table~\ref{table:ta}). 
 
\begin{figure}[t]
\centering
 \centering
  \includegraphics[width=0.49\linewidth]{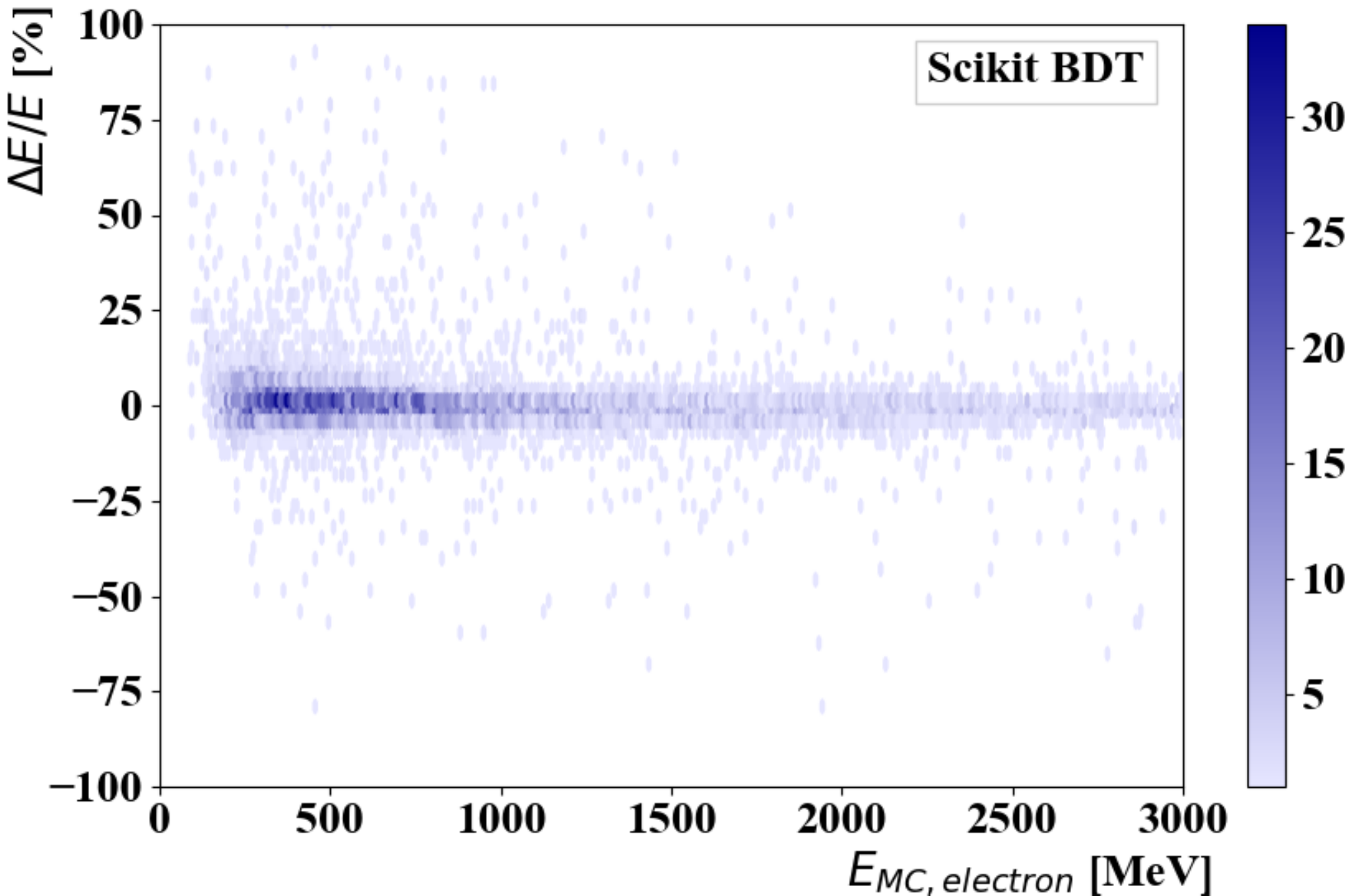}
 \centering
  \includegraphics[width=0.49\linewidth]{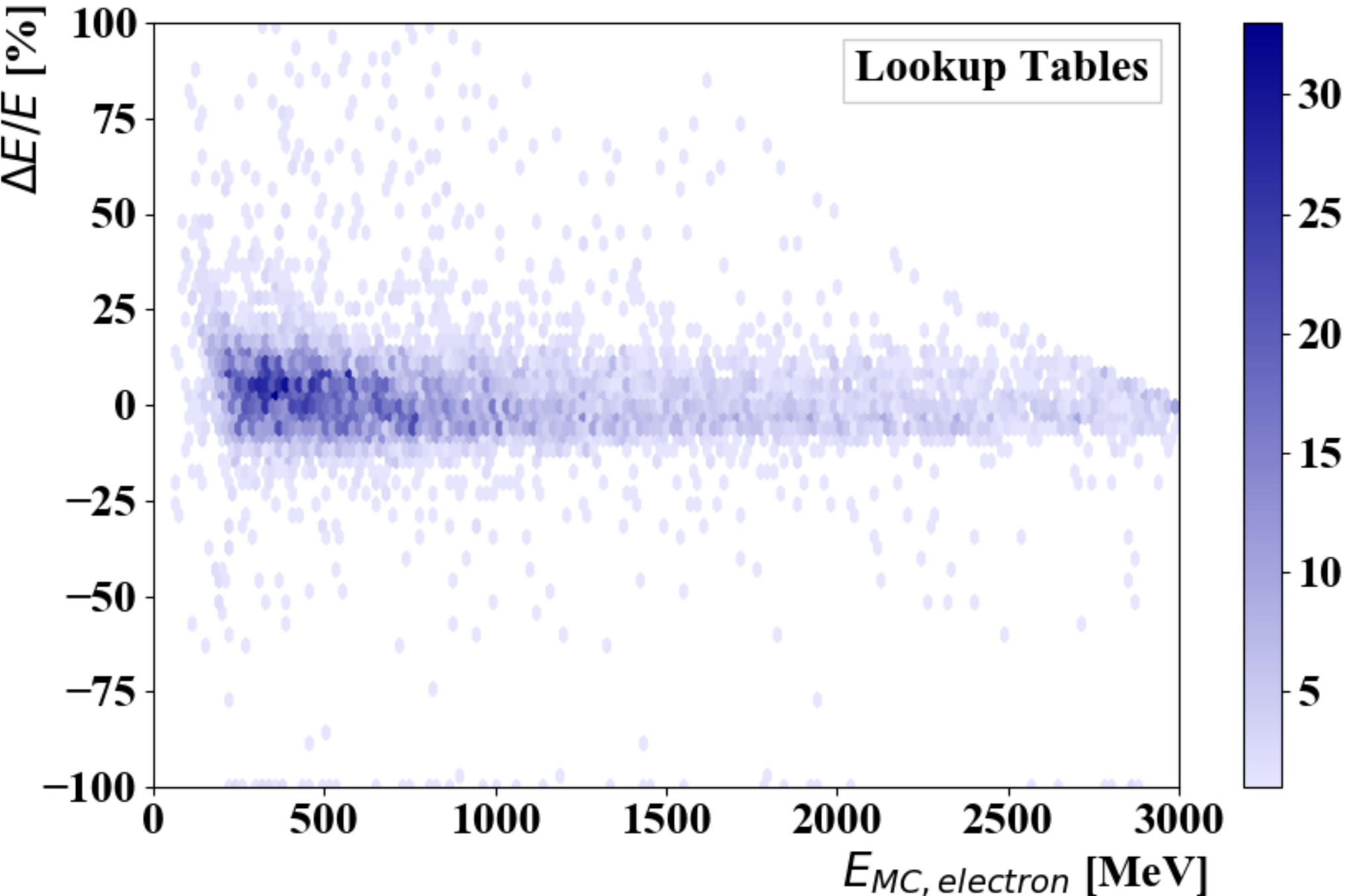}
\caption{The ${\Delta E}/{E}$ distribution as a function of the MC electron energy for the BDT from Scikit-Learn (left) and the looukup tables (right)  for events in the default fiducial volume. The colour code represents the density of events for each bin.}
\label{fig:resINfidel}
\end{figure}

\begin{figure}[t]
\centering
 \includegraphics[width=0.49\linewidth]{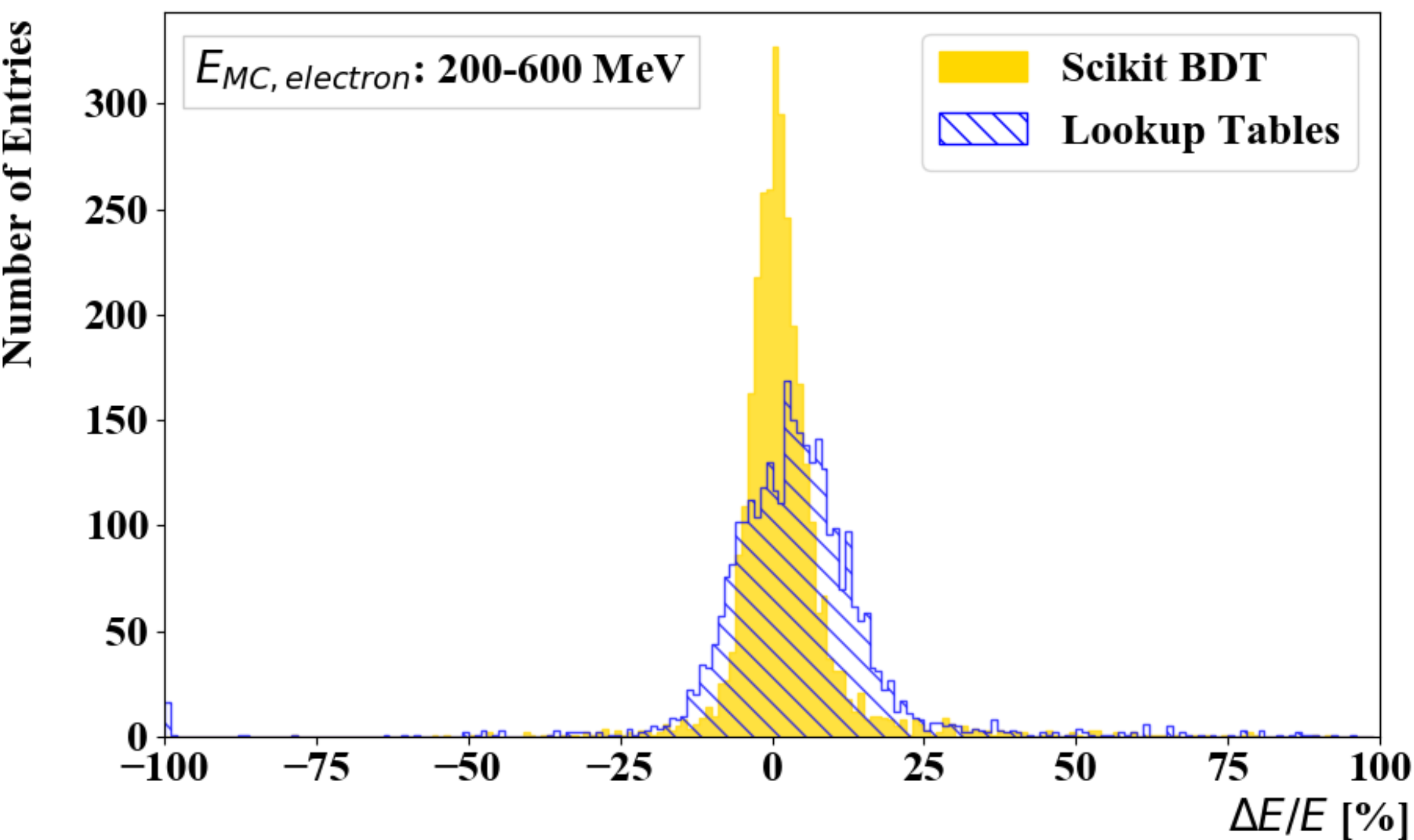}
 \centering
 \includegraphics[width=0.49\linewidth]{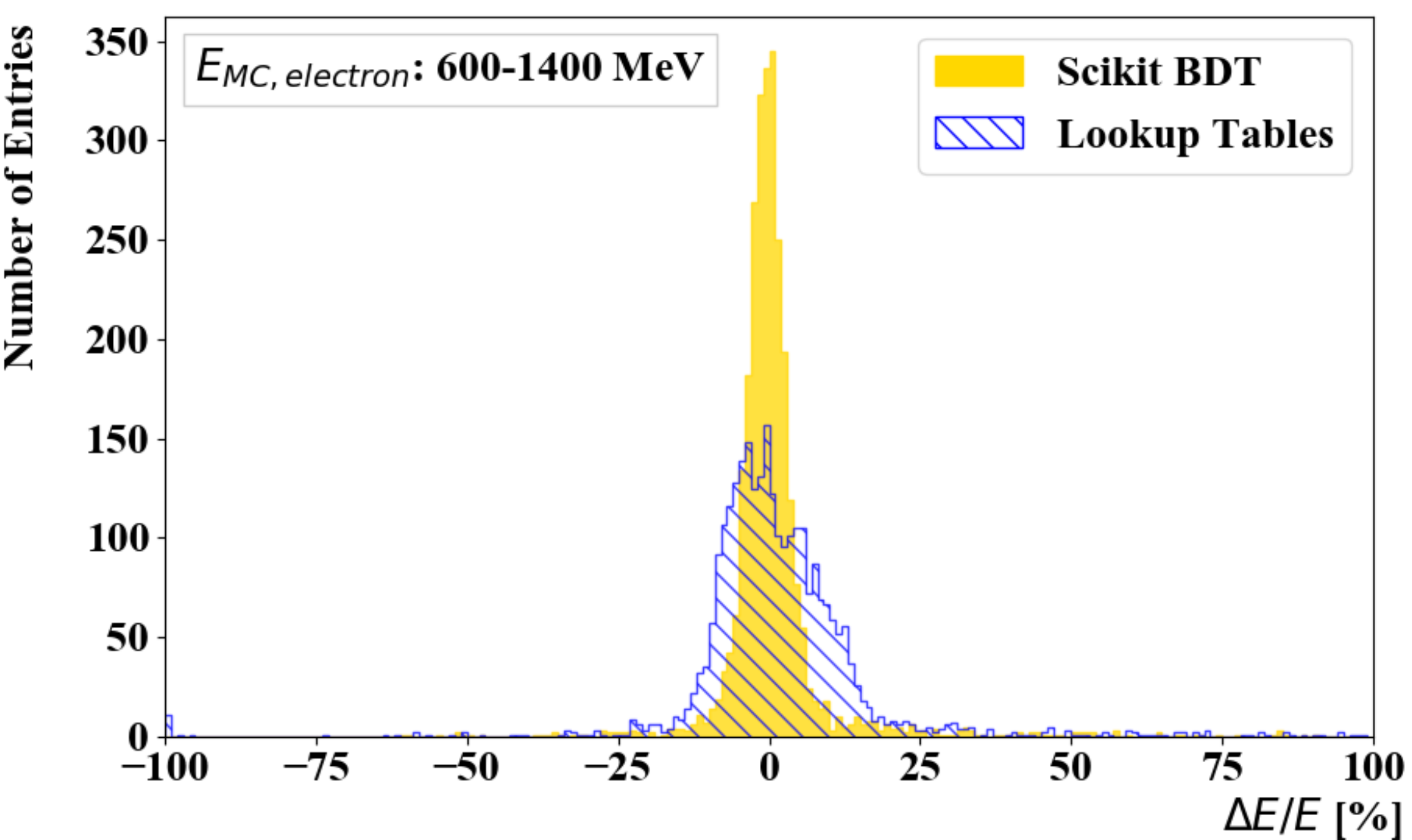}
  \caption{The distribution of ${\Delta E}/{E}$ for events in the default fiducial volume and electron energies from 200-600 MeV (left plot) and from 600-1400 MeV (right plot) for the BDT from Scikit-Learn and the lookup tables.}
  \label{fig:resnueINFID}
\end{figure}

The reconstruction of the electron energy in charged-current electron neutrino interactions has also been studied using a Scikit-Learn BDT trained on an electron sample. The results are shown in Figure~\ref{fig:resINfidel} (left) for events in the default fiducial volume. These can be compared with the corresponding results using the lookup tables (Figure~\ref{fig:resINfidel} (right)) for the same sample of events.  The BDT leads to better energy reconstruction for electrons for the whole energy range compared to lookup
tables. This can also be seen in Figure~\ref{fig:resnueINFID} which shows the distributions of the fractional energy residuals for both approaches and for electron energies between 200-600 MeV and 600-1400 MeV. The results of the fit procedure for electrons in the default fiducial volume are reported in Table~\ref{table:electronINfid}. The Scikit-Learn BDT leads to better energy resolution compared to the lookup table approach for both electron energy ranges.

\begin{table}[t]
\caption{Fitted energy resolution parameters for electron neutrino interactions within the fiducial volume. The quoted uncertainties are statistical.}
\vspace{\baselineskip}
 \centering
   \begin{tabular}[b]{|c | c| c | c | c| c|}
     \hline
     \multirow{2}{*}{\textbf{Method}} & \multirow{2}{*}{\textbf{Energy Range [MeV]}}& \multicolumn{4}{c}{\textbf{${\Delta E}/{E}~\%$}} \vline \\  \cline{3-6}
     & & \textbf{$\mu$}  & \textbf{$\sigma_R$} & \textbf{$\sigma_L$} & Tail$~\%$ \\  \hline
     Scikit-Learn BDT & 200-600 & \phantom{$-$}0.22$\pm$0.24  &4.52$\pm$0.19  &3.67$\pm$0.15 &9.24 \\
           & 600-1400 & {$-$}0.64$\pm$0.14  &3.45$\pm$0.10  &3.15$\pm$0.10 &8.42 \\
     Lookup Tables & 200-600 & \phantom{$-$}4.84$\pm$0.49 & 5.45$\pm$0.96 & 10.15$\pm$0.58  &7.23 \\
           &600-1400 & {$-$}4.69$\pm$0.37 & 11.32$\pm$0.69 & 3.95$\pm$0.45  &6.23 \\
     \hline
   \end{tabular}
\label{table:electronINfid}
\end{table}

The effect of the extended fiducial volume has also been studied for electrons. The fiducial volume extension leads to comparable energy resolution for the Scikit-Learn BDT,
while it increases the number of reconstructed events by $\sim$~9~\% for electrons with energy between 200-600~MeV and $\sim$~10~\% for electron energies in the range 600-1400~MeV.
The energy reconstruction for electrons leads to better results than
for muons for the high energy range (600-1400 MeV). This is because electrons with energies above 1~GeV have a path length of approximately 2-2.5 m and the majority of them will not escape the detector volume. However, muons at this energy range have a path length of approximately 2.5-10 m and are thus more likely to escape the instrumented volume~\cite{path}. As a result in the electron case we do not have large over- (or under-) estimations of the energy resulting in improved resolution compared to the muon case.
 
 \section{Discussion}
 \label{discussion}
 The Scikit-Learn BDT is fast to train (309 seconds on a  2.90 GHz dual core Intel Core i5 processor) and easy to implement in different detector configurations. The results achieved are comparable to the ones reported by the Super-Kamiokande and Hyper-Kamiokande experiments using likelihood function-based techniques (fiTQun)~\cite{fitqun}. The results for  
 Super-Kamiokande with 20$\%$ photocove-rage (SK-II ), 40$\%$ photocoverage (SK-IV) and the expected performance of Hyper-Kamiokande with 20$\%$ photocoverage are summarised in~\cite{hyperk}. Muons and electrons with 500 MeV/$c$ were generated with a fixed vertex (at the centre of the tank) and direction (toward the barrel of the tank). The fiTQun reconstruction was applied. The momentum resolution reported is 3.6\%, 2.3\% and 2.6\% (5.6\%, 4.4\% and 4.0 \%) for muons (electrons) in SK-II, SK-IV and Hyper-Kamiokande respectively~\cite{hyperk}. These results can be compared with the energy resolution achieved by Scikit-Learn BDT of $\sigma_R$=3.42$\pm$0.05\%, $\sigma_L$=2.90$\pm$0.05\% ($\sigma_R$=4.05$\pm$0.31\%, $\sigma_L$=3.62$\pm$0.28\%) for muons (electrons) with energies between 400-600 MeV within the default fiducial volume.   
The larger Super-Kamiokande and Hyper-Kamiokande fiducial volumes, the different sample selections, the different electronics and the better direction and vertex reconstruction for these detectors do not allow for a direct comparison of the two methods. However, at a first approximation it is shown that using a BDT leads to comparable results to techniques that require the construction of detailed likelihood functions. 

\section{Robustness tests}
\label{robustness} 
When using a multivariate technique there is a concern that the results are dependent on the accuracy of the modelling of the input variables in the training samples.
In order to address this question, the performance of the Scikit-Learn BDT was re-evaluated using samples of simulated events varying the quantum efficiency of the PMTs. This variable was chosen because the number of photoelectrons is the most important input to the method.

\begin{figure}[t]
\centering
\includegraphics[scale=0.35]{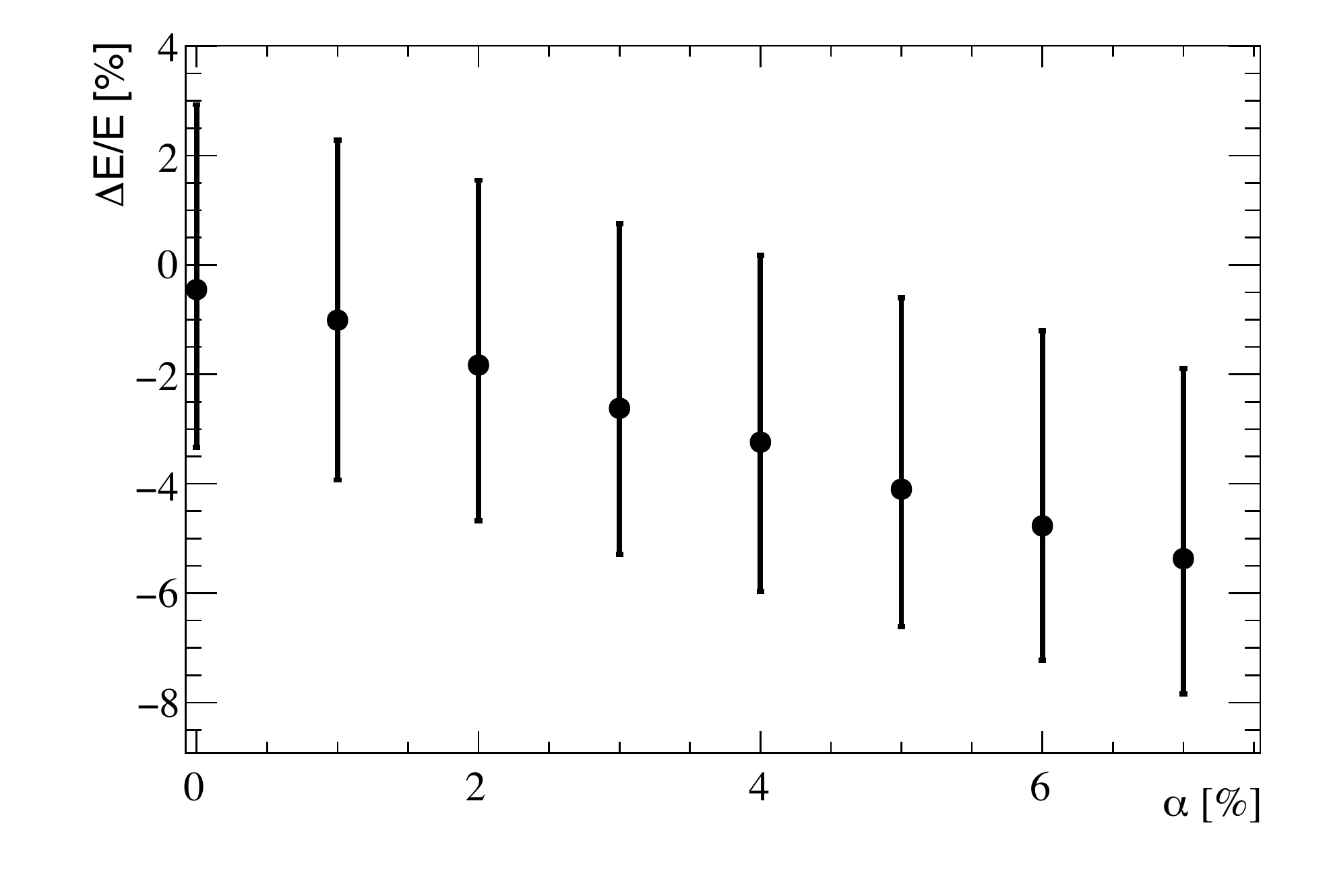}
\caption{Variation of $\mu$ versus the inefficiency parameter $\alpha$ of the testing sample for muon neutrino interactions within the fiducial volume with muon energies between 200-600 MeV.
The bars drawn above and below the mean indicate $\sigma_{R}$ and $\sigma_{L}$ of the bifurcated Gaussian, respectively.}
\label{fig:recoAll}
\end{figure}

For this study, charged-current muon neutrino events with true muon energies between 200-600 MeV within the fiducial volume were selected.
The bifurcated Gaussian fit, described in Section~\ref{results}, was used to determine the value of $\mu$, $\sigma_R$ and $\sigma_L$.
The BDT was then tested using samples with a different quantum efficiency, given by  $\epsilon = (100-\alpha)* \epsilon_0 /100 $, 
where $\epsilon_0$ is the quantum efficiency of the training sample and $\alpha$ is varied for each sample.
Figure~\ref{fig:recoAll} shows that there is a linear increase in $\mu$ with increasing inefficiency but no significant change in the width of the distributions.
The other methods described in this paper showed the same linear increase with increasing inefficiency when tested. An additional test was carried out where each PMT was assigned a quantum efficiency value sampled from a Gaussian distribution with a mean of 21\% and a sigma of 0.5. In these tests the Scikit-Learn BDT was robust and performed comparably to its performance on the sample with unaltered quantum efficiency. 

The dark noise rate also affects the number of detected photoelectrons and so this parameter was also varied in each testing sample.  Dark noise rates up to 3~kHz per PMT, which are the reference dark noise rates for current E61 and Hyper-K photomultiplier studies, were tested but with only a small difference in the fit parameters visible,
as shown in Table~\ref{table:darknoiseTable}.
We conclude that the Scikit-Learn BDT performance is robust and no more difficult to calibrate than the lookup tables.

\begin{table}[t]
 \caption{ Fitted energy resolution parameters for muon neutrino interactions within the fiducial volume with the muon energy in the range 200 to 600 MeV for the Scikit-Learn BDT. The quoted uncertainties are statistical.}
\vspace{\baselineskip}
 \centering
   \begin{tabular}[b]{|c | c | c | c| c|}
     \hline
     \multirow{2}{*}{\textbf{Dark~Noise}} & \multicolumn{4}{c}{\textbf{${\Delta E}/{E}~\%$}} \vline \\  \cline{2-5}
      & \textbf{$\mu$}  & \textbf{$\sigma_R$} & \textbf{$\sigma_L$} & Tail$~ \%$ \\  \hline
     0 kHz & $-$0.73$\pm$0.18 &3.67$\pm$0.13 & 2.86$\pm$0.12& 3.60\\
     3 kHz & $-$0.14$\pm$0.21 & 3.04$\pm$0.14& 2.95$\pm$0.14& 5.00\\
     \hline
   \end{tabular}
\label{table:darknoiseTable}
\end{table}

\section{Conclusions}
\label{conclusions}
We have discussed the use of machine learning-based methods for charged-lepton energy reconstruction in neutrino interactions in water Cherenkov detectors. Several different machine learning algorithms have been implemented using standard software packages and all have demonstrated significantly improved energy resolution compared to a traditional lookup-table based approach. The best performance is achieved with the gradient BDT from the Scikit-Learn package. The Scikit-Learn BDT is fast to build and train (309 seconds on a  2.90 GHz dual core Intel Core i5 processor) and easy to implement in different detector configurations. This method gives better energy resolution and reduced tails compared to the NN and the lookup-table based approach adopted during the TITUS design. Scikit-Learn based BDT has the additional advantage of allowing the optimisation of the adjustable parameters using a rectangular grid search which is not available at the TMVA package. The improved resolution will allow a larger fiducial volume to be defined for the same size of detector, providing a more cost-effective approach for detector design. The robustness of the BDT algorithm was tested and the results indicate stable
performance as a function of the PMT quantum efficiency and dark-noise rate. This method has been adapted for use in ANNIE Phase II~\cite{ANNIEPhase2} and first results are encouraging. It is also easily adaptable for use in E61 where a direct comparison with a likelihood-function based technique, fiTQun~\cite{fitqun}, can be performed. Such techniques give excellent performance but require the construction of detailed likelihood functions and thus are time-consuming (regarding the human time) to optimise for each detector geometry. In contrast, the machine-learning approaches discussed here are easily implemented and fast to be trained while also giving good performance. More details on the code can be found in Ref.~\cite{github}.


\acknowledgments

We acknowledge support from STFC (United Kingdom), in particular grant ST/K004646/1, and the TITUS collaboration for the use of their simulation and reconstruction software.

\end{document}